\documentclass[12pt]{article}
\usepackage{graphicx}
\usepackage{amsmath,amssymb,amsfonts}

\renewcommand{\theequation}
{\arabic{section}.\arabic{equation}}

\makeatletter
\def\eqnarray{ \stepcounter{equation} \let\@currentlabel=\theequation
 \global\@eqnswtrue
 \global\@eqcnt\z@
 \tabskip\@centering
 \let\\=\@eqncr
 $$\halign to \displaywidth\bgroup\@eqnsel\hskip\@centering
 $\displaystyle\tabskip\z@{##}$&\global\@eqcnt\@ne
 \hfil$\displaystyle{{}##{}}$\hfil
 &\global\@eqcnt\tw@$\displaystyle\tabskip\z@{##}$\hfil
 \tabskip\@centering&\llap{##}\tabskip\z@\cr}
\makeatother

\makeatletter
\def\@arrayacol{\edef\@preamble{\@preamble \hskip .5\arraycolsep}}
\def\array{\let\@acol\@arrayacol \let\@classz\@arrayclassz
\let\@classiv\@arrayclassiv \let\\\@arraycr\def\@halignto{}\@tabarray}
\makeatother

\makeatletter
\newcounter{subeqncnt}
\def\thesubeqncnt{\alph{subeqncnt}}
\def\subequations{\begingroup%
   \stepcounter{equation}\edef\@tempa{\theequation}%
   \let\c@equation\c@subeqncnt\c@subeqncnt\z@
   \edef\theequation{\@tempa\noexpand\thesubeqncnt}}

\makeatother

\newcommand{\be}{\begin{equation}}
\newcommand{\ee}{\end{equation}}

\newcommand{\beqa}{\begin{eqnarray}}
\newcommand{\eeqa}{\end{eqnarray}}
\newcommand{\nn}{\nonumber}

\begin{document}

\setlength{\baselineskip}{7mm}
\begin{titlepage}
\begin{flushright}

{\tt NRCPS-HE-1-2015\\
SigmaPhi2014, July 2014\\
Rhodes , Greece} \\

\end{flushright}

\begin{center}
{\Large ~\\{\it  The Gonihedric Paradigm \\
\vspace{0,6cm}
Extensions of the Ising Model
}

}

\vspace{1cm}
 
{\sl George Savvidy

\bigskip
\centerline{${}$ \sl Institute of Nuclear and Particle Physics}
\centerline{${}$ \sl Demokritos National Research Center} 
\centerline{${}$ \sl Ag. Paraskevi,  Athens, Greece}
\bigskip

}
 
\end{center}
\vspace{30pt}

\centerline{{\bf Abstract}}

We  suggest  a generalization of the Feynman path integral to an integral over random surfaces. The proposed action is proportional to the linear size of the random surfaces and  is called gonihedric. The convergence and the properties of the partition function are analysed.
The model can also be formulated as a spin system with identical partition function.
The  spin system represents a generalisation of the Ising model with ferromagnetic,
 antiferromagnetic and quartic interactions. Higher symmetry of the model
 allows to construct dual spin systems in three and four dimensions.
 In three dimensions the transfer matrix describes the propagation of closed loops
 and we found its exact spectrum. It is a unique exact solution of the
 tree-dimensional statistical spin system. In three and four dimensions the system exhibits the second order phase transitions.
The gonihedric spin systems have  exponentially degenerated vacuum states
separated by the potential barriers and can be used as a storage of binary information.

\vspace{12pt}

\noindent

\end{titlepage}

\pagestyle{plain}
 
\pagestyle{plain}

\section{\it Extension of Feynman Path Integral }

Feynman path integral over  trajectories describes quantum-mechanical behaviour  of point-like
particles, and it is an important problem to extend the path integral to an integral which describes
a quantum-mechanical motion of strings.  A string is a one dimensional extended object which
 moves  through the space-time.  As string moves through the space-time it sweeps out
 a two-dimensional surface, and in order to describe its quantum-mechanical behavior
one should define an appropriate action and  the corresponding
functional integral over two-dimensional surfaces.

\begin{figure}
\includegraphics[width=10cm]{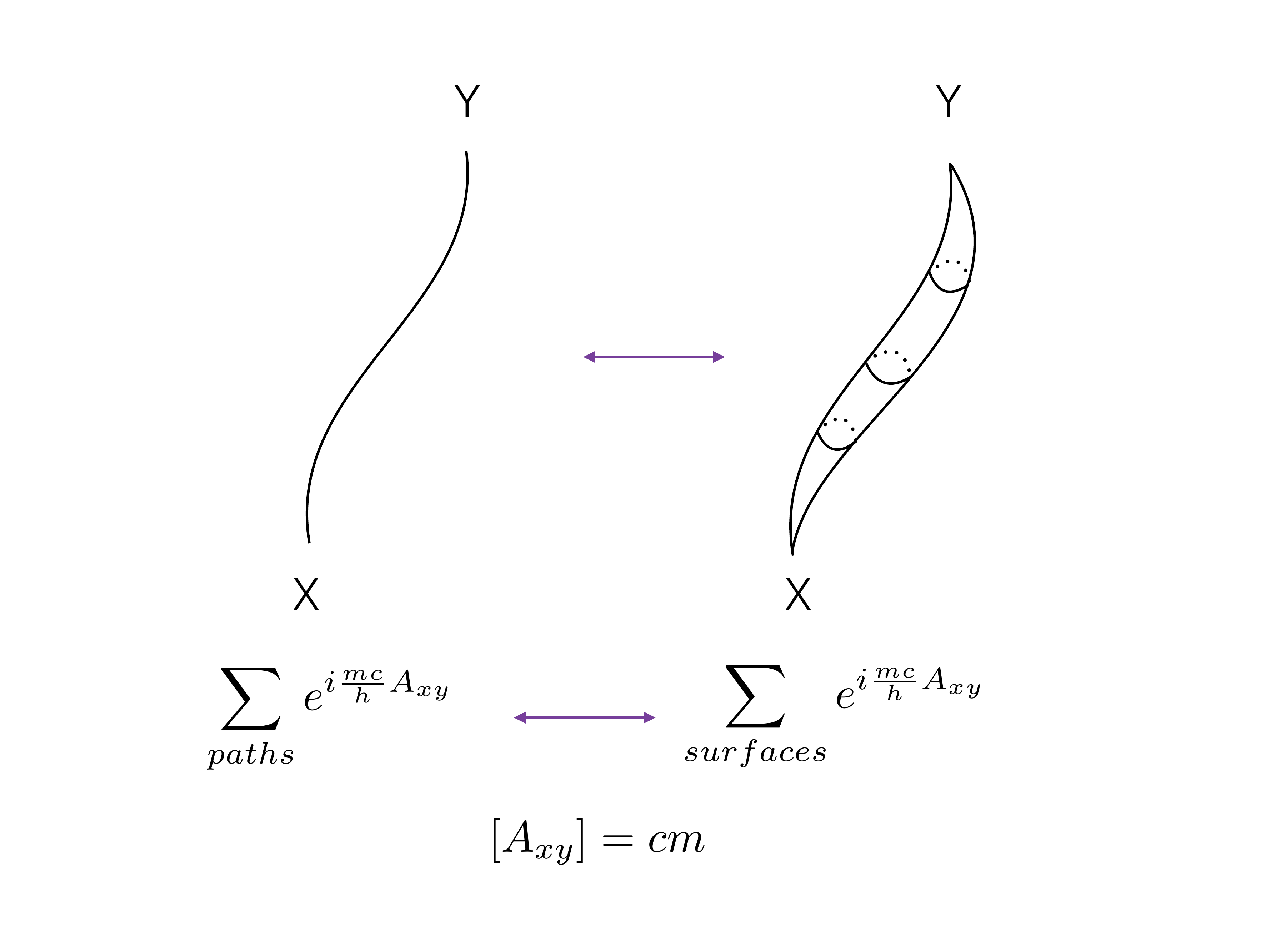}
\centering
\caption{On the left hand side of the figure there is a trajectory of a structureless point particle with an action which is proportional to the length of its world line. On the right there is a world sheet surface which is  swept by a propagating closed string. It is required that the string action $A_{xy}$  should be proportional to the linear size of the space-time surface, measuring it in terms of its length, similar to the action of a point-like particle. This is a natural requirement because when a string collapses to a point its world sheet will degenerate into a world line and both actions will coincide.}
\label{fig1}
\end{figure}

In string theory the action is defined by using Nambu-Goto  {\it area action} \cite{223,Goto:1971ce}.
The area action suffers from spike instabilities \cite{Ambjorn:1985az}, because the
zero-area spikes can easily grow  on a surface. Indeed, the spikes have zero area,
and there is no suppression of the spike fluctuations in the functional integral.
Different  modifications of the area action have
been suggested in the literature  to cure these instabilities,
which are based on the addition of extrinsic curvature
terms \cite{Polyakov:1986cs} to the area action.

\begin{figure}
\includegraphics[width=6cm]{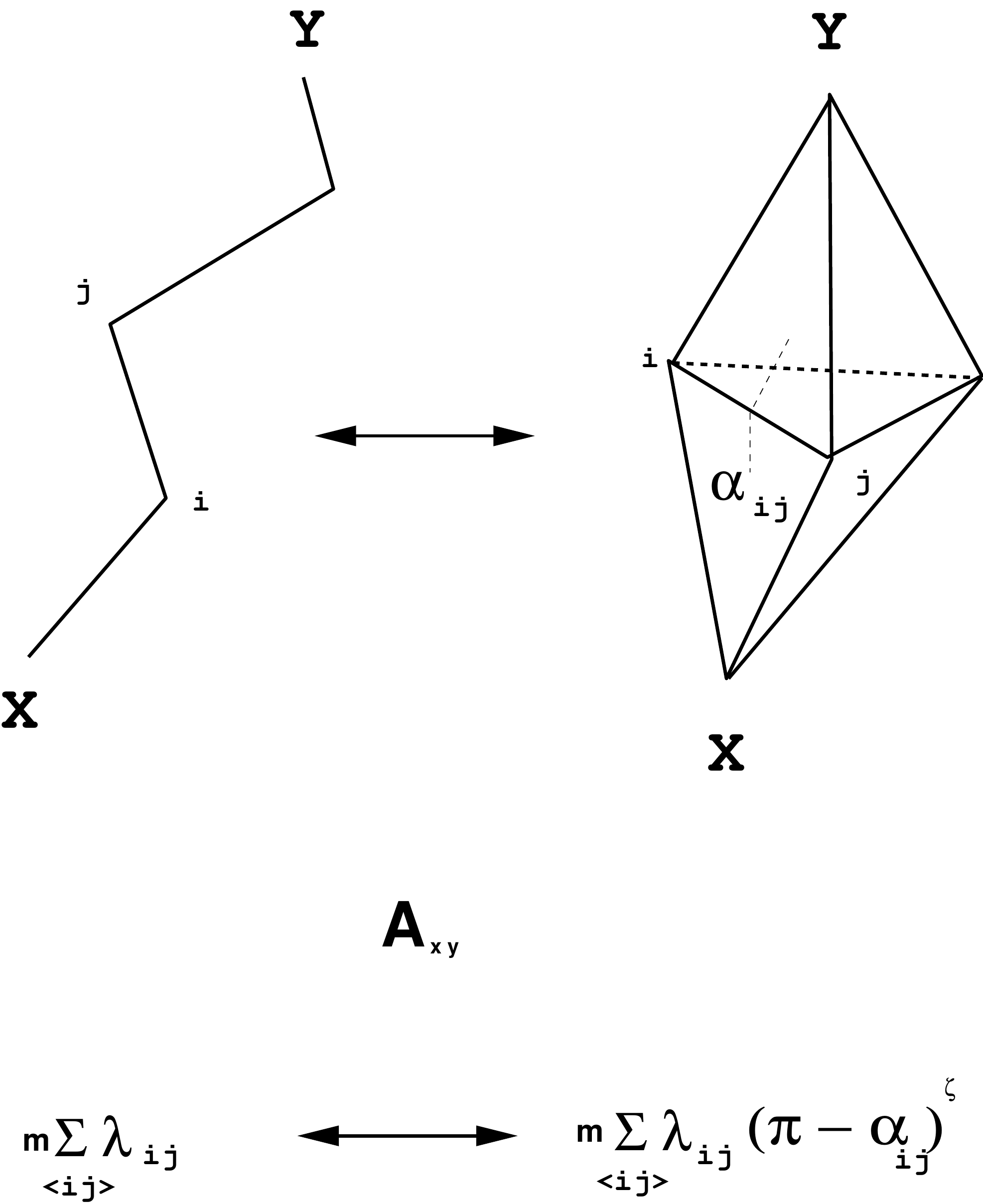}
\centering
\caption{On the left there is a discrete trajectory with an  action which is
proportional to the sum of the lengths of its edges. On the right there
is a discretized surface and the
action is  a sum of the lengths of its edges multiplied by the corresponding deficit angles.}
\label{fig2}
\end{figure}

The alternative principle to cure surface instabilities was
put forward in \cite{1,2,3}.
In its essence there is a new requirement which should be
imposed on the string action.  The string action
should be defined in such a way that when a string shrinks to a point-like object
its action should reduce to an action of point-like particle \cite{1}.
In other words, when a surface shrinks to a space-time  curve,
its action should reduce to the length of the curve (see Fig.~\ref{fig1} ).
It is almost obvious that now the spikes cannot easily grow on a surface because
in the functional integral such fluctuations will be  suppressed exponentially
$e^{-m \lambda_{spike}}$, where $\lambda_{spike}$ is the length of the
spike.

One can consider smooth surfaces, as well as discretized random surfaces
which are represented by polyhedral surfaces build from triangles (see Fig.~\ref{fig2}).
For smooth surfaces the proposed action has the form
\cite{Savvidy:2001pm,Savvidy:2003dv,Savvidy:2003fx,Savvidy:2005fe,Mourad:2005rt}
\begin{equation}\label{continumaction}
 A(M)=m \int d^{2}\zeta
  \sqrt{g}\sqrt{ \left( \Delta(g) X^{\mu} \right)^{2}} ~~~~\rightarrow  ~~~~
  m \int  d s,
\end{equation}
where $ g_{ab}=\partial_{a} X^{\mu} \partial_{b} X^{\mu} $ is the induced metric and
$\Delta(g)$ is a Laplace operator.  For discretized surfaces the action is defined
as a sum over links  and deficit angles \cite{1,2,3}:
\begin{equation}\label{energyfunction}
A(M)= m \sum_{<ij>} \lambda_{ij} \vert \pi - \alpha_{ij} \vert ~~~~~\rightarrow ~~~~~~~~~m \sum _{<ij>} \lambda_{ij}.
\end{equation}
The action involves
the products of edge length  $\lambda_{ij}=\vert x_i -x_j \vert$ times the corresponding
deficit angle $\vert \pi - \alpha_{ij} \vert$, and it was suggested
to call that action - gonihedric from two hellenic words $\gamma \omega\nu \iota' \alpha$ -
the angle and $\epsilon' \delta\rho\alpha$  - the side.
The corresponding partition function can be represented in the
form\footnote{The partition function (\ref{gonihedric})
is defined as an integral over all surface vertices
of a given triangulation and  a summation over all topologically different
triangulations\cite{2} - \cite{5}. We proved that the contribution of a given
triangulation to the partition function is finite and found the explicit form for the
upper bound\cite{4}. The question of the convergence of the sum over all triangulations of the
full partition function remains open\cite{Durhuus:1992np,5}.}
\begin{equation}\label{gonihedric}
Z(\beta) = \sum_N \int d^D x_1....d^D x_N \exp{(-\beta m \sum_{<ij>}  \vert x_i -x_j \vert
\vert \pi - \alpha_{ij}  \vert  ) }.
\end{equation}
The general arguments, based on the Minkowski inequality \cite{1},
show that the maximal contribution to the partition function comes from
the surfaces close to a sphere.  The convergence of the partition function was   proven
rigorously in \cite{4,Durhuus:1992np,5} .
Because the action is proportional to the "perimeter" of the surface, and  not to its area,
the classical string tension is  equal to zero and the model can equally well be called a
model of tensionless strings
\cite{Savvidy:2001pm,Savvidy:2003dv,Savvidy:2003fx,Savvidy:2005fe}.

It is  important to study the phase structure of the system
(\ref{gonihedric}) in order to identify  quantum field theory to
which it is   equivalent near its critical point \cite{Wilson:1974,Wilson:1973jj,Cardy:1989da}.
At low temperature  the Wegner-Wilson loop correlation functions
have perimeter behavior.  At high temperature the   fluctuations
of the surface  induce the area behavior
signaling that there is a confinement - deconfinement
(or area - perimeter)  phase transition \cite{1,2,Savvidy:2001pm}.
In this model a nonzero string tension is
entirely generated by quantum (thermodynamical)  fluctuations
 \cite{2,Savvidy:2001pm,3}.
\begin{figure}
\includegraphics[width=6cm]{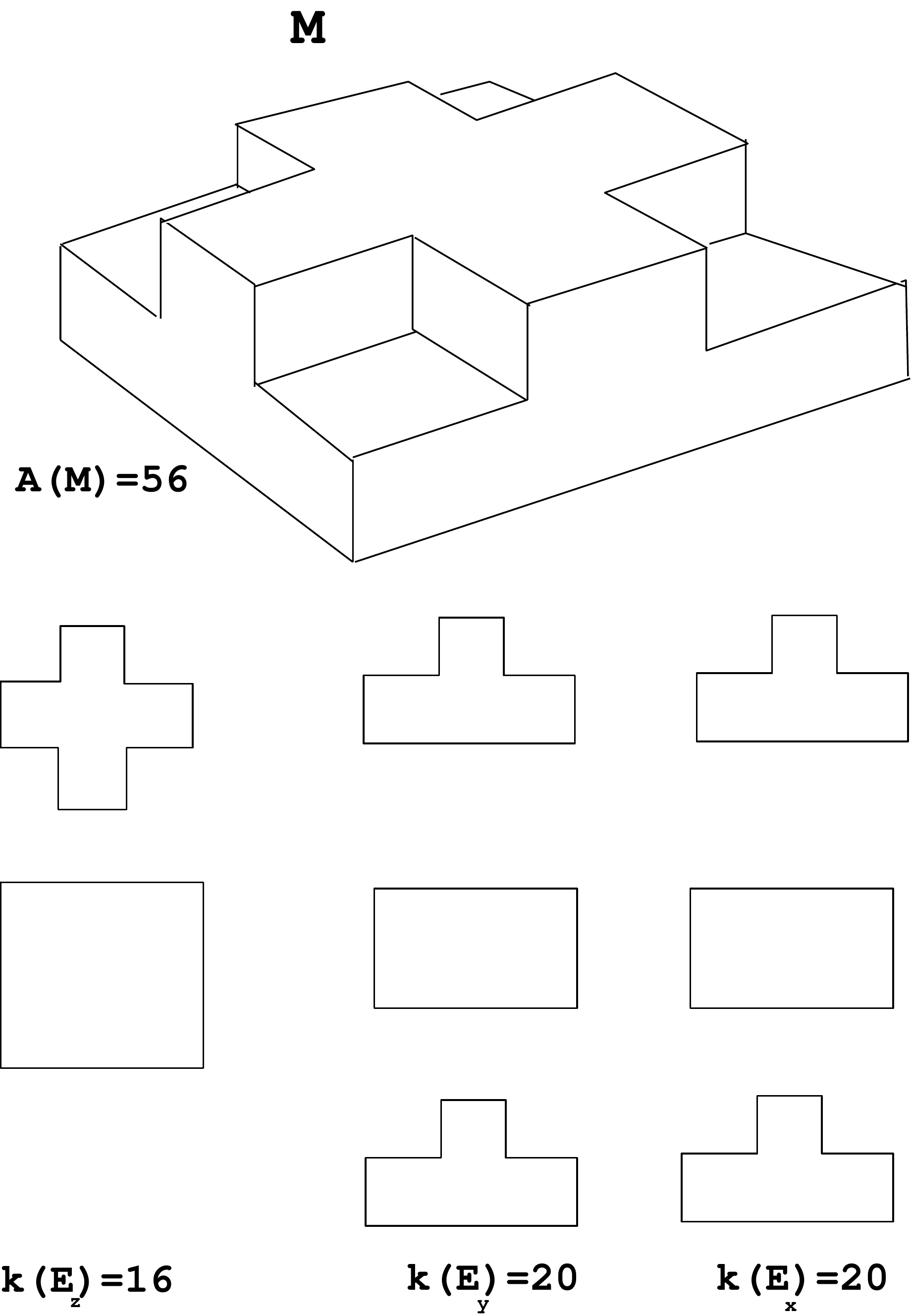}
\centering
\caption{ On the top is an example of a surface M on hyper cubic lattice.
The set of planes
$\{ E_{x} \}, \{ E_{y} \}, \{ E_{z} \} $ perpendicular to $x, y,z $ axis
intersect a given surface $M$ in the middle of the links.
On each of these planes we shall have an image of the  surface
$M$. Every such image is represented by closed polygon-loops
$ P(E)$ appearing in the intersection of
the plane with surface $M$.
The energy of the surface $M$ is equal to the
sum of the total curvature $k(E)$ on all these planes, as it is
given by the formula (\ref{latcurv}).
For the surface on the picture this sum is equal to
$A(M) = 16+20+20= 56$ times $ a {\pi \over 2}$.
 As we shall see, the
energy of the surface M can be recovered by using images of the surface only on one
sequence of  parallel  planes (\ref{commonact}). On  Fig.~\ref{fig4} this is demonstrated by
using only images of the surface M on $\{E_z\}$ planes. }
\label{fig3}
\end{figure}

\section{\it Extensions of the Ising and Wegner Models}

The  model of gonihedric random surfaces was formulated as embedding
of random surfaces into the Euclidean space \cite{1,2,3}. It can also be formulated as
a model of random surfaces embedded into the hyper-cubic lattice \cite{6,7,8,9,10}
(see Fig.~ \ref{fig3}).
The advantage
of the lattice formulation consists in the fact that one can construct a spin system which is
a generalisation of the Ising model with ferromagnetic, antiferromagnetic and quartic
interactions so that its interface energy is equal to the gonihedric energy (\ref{energyfunction}).

On the lattice a closed surface $M$ can be considered as a collection
of plaquettes the edges of which are glued together pairwise.
The surface is considered as a connected, orientable
surface with given topology, and it is assumed that
self-intersections of the surface  produce additional contributions to  the
energy functional proportional to the length of the intersections.
On the lattice the lengths of the elementary edges
$\lambda_{ij}= \vert x_{i} - x_{j} \vert $ are equal to the lattice constant $a$ and
the angles between plaquettes are either $0$ or $ \pi /2$.
The gonihedric energy functional (\ref{energyfunction}) on the hyper-cubic lattice
therefore takes the following  form:
\be\label{energy}
A(M) = (n_2  + 4  k  n_4)~ a ~{\pi \over 2}
\ee
where $n_2$ is the total number of edges on which two
plaquettes intersect at right angle, $n_4$ is the total number of edges with four intersecting plaquettes
and $k$ is {\it the self-intersection coupling constant.  It describes the intensity
of string interactions: string can split into two strings and merge with other strings
(see Fig. \ref{fig7}).}
The equivalent Hamiltonian of the spin system on three-dimensional lattice has the form
\cite{6,7,8}
 \be\label{spinsystem}
 H^{3d}_{gonihedric}=- 2k \sum_{\vec{r},\vec{\alpha}}
\sigma_{\vec{r}} \sigma_{\vec{r}+\vec{\alpha}} + \frac{k}{2}
\sum_{\vec{r},\vec{\alpha},\vec{\beta}} \sigma_{\vec{r}}
\sigma_{\vec{r}+\vec{\alpha} +\vec{\beta}} -  \frac{1-k}{2}
\sum_{\vec{r},\vec{\alpha},\vec{\beta}} \sigma_{\vec{r}}
\sigma_{\vec{r}+\vec{\alpha}}
\sigma_{\vec{r}+\vec{\alpha}+\vec{\beta}}
\sigma_{\vec{r}+\vec{\beta}},
\ee
where the vector $\vec \alpha$ runs over the unit vectors parallel to
the axes. Similarly, the sum over $\vec \alpha$ and $\vec
\beta$ runs over different pairs of such vectors.
The Hamiltonian represents a magnetic system with competing
interaction and specially adjusted coupling constants
$ J_{ferr} = 4J_{antiferr}=2k$.
The Hamiltonian (\ref{spinsystem}) contains the usual Ising ferromagnet
$$
H_{Ising}^{3d}= - J \sum_{\vec{r},\vec{\alpha}} \sigma_{\vec{r}}
\sigma_{\vec{r}+\vec{\alpha}}
$$
with additional diagonal antiferromagnetic interaction and
quartic spin interaction which regulates the intensity of the interaction
at the self-intersection edges. When $k=0$ the surfaces can freely
intersect, at $k=\infty$ the surfaces  are strongly self-avoiding  \cite{7,8}.

The partition functions of both systems (\ref{energy})  and (\ref{spinsystem})
are identical to each other:
\be\label{equalpartition}
Z(\beta) = \sum_{\{M\}} e^{-\beta~A(M)} ~
= ~\sum_{\{\sigma\}} e^{-\beta H_{gonihedric}(\sigma)}.
\ee
In the first case it is a sum over all two-dimensional surfaces of the type
described above with energy functional $A(M)$,  in the
second case it is a sum over all spin configurations.
This spin system has very high symmetry because one can flip the spins  on any flat
hypersurface without changing the energy of the system. {\it The rate of degeneracy of
the vacuum state depends on the self-intersection coupling constant $k$ \cite{8,9}.
If $k \neq 0$, the degeneracy of the vacuum state is equal to $3\cdot 2^N$ for
the lattice of the size $N^3$, because one can flip spins on any set of parallel
planes \cite{8,9} }.

A similar construction can be performed in four dimensions \cite{6,7}.
On a four-dimensional lattice  a two-dimensional
closed surface can have self-intersections of different orders because at a
given edge one can have self-intersections of four or six plaquttes.
The energy which is ascribed to self-intersections essentially
depends on a configuration of plaquettes in the intersection.
There are two topologically different
configurations of plaquettes with four intersecting plaquettes and only
one with six intersecting plaquettes. The corresponding spin
system is locally  gauge invariant \cite{7}:
\beqa \label{gonih}
H^{4d}_{gonihedric}= -\frac{5\kappa-1}{g^{2}} \sum_{\{plaq\}}
(\sigma\sigma\sigma\sigma) + \frac{\kappa}{4g^{2}}
\sum_{\{right~angle~plaq\}}
(\sigma\sigma\sigma\sigma_{\alpha})^{rt}
(\sigma_{\alpha}\sigma\sigma\sigma)\nn\\
-\frac{1-\kappa}{8g^{2}}
\sum_{\{triples~of~right~angle~plaq\}}
(\sigma\sigma\sigma\sigma_{\alpha})^{rt}
(\sigma_{\alpha}\sigma\sigma\sigma_{\beta})^{rt}
(\sigma_{\beta}\sigma\sigma\sigma)  .
\eeqa
The total energy of the surface in this  case is
\be\label{energyk}
A(M)= (n_{2} + 4k \bar{n}_{4} + (6k-1) \bar{\bar{n}}_{4} + 12k n_{6})~ a ~{\pi \over 2},
\ee
where  $n_{2}$ is the total number of edges, where two
plaquettes intersect at the right angle, $\bar{n}_{4}$ and
$\bar{\bar{n}}_{4}$ edges with
intersection of four plaquettes and $n_{6}$ with six plaquettes.
The partition functions of both four-dimensional systems (\ref{gonih}) and (\ref{energyk}) are
identical to each other,  as in (\ref{equalpartition}).
Thus a sum over all two-dimensional surfaces of the type
described above with energy functional $A(M)$, embedded into a
four-dimensional lattice is identical to a
sum over all spin configurations.

The spin systems described above can be studied by powerful analytical
methods \cite{9,10,27,28,11,12,13},  as well  as by Monte-Carlo simulations \cite{14,15,16,17,26}.

\begin{figure}
\includegraphics[width=7cm]{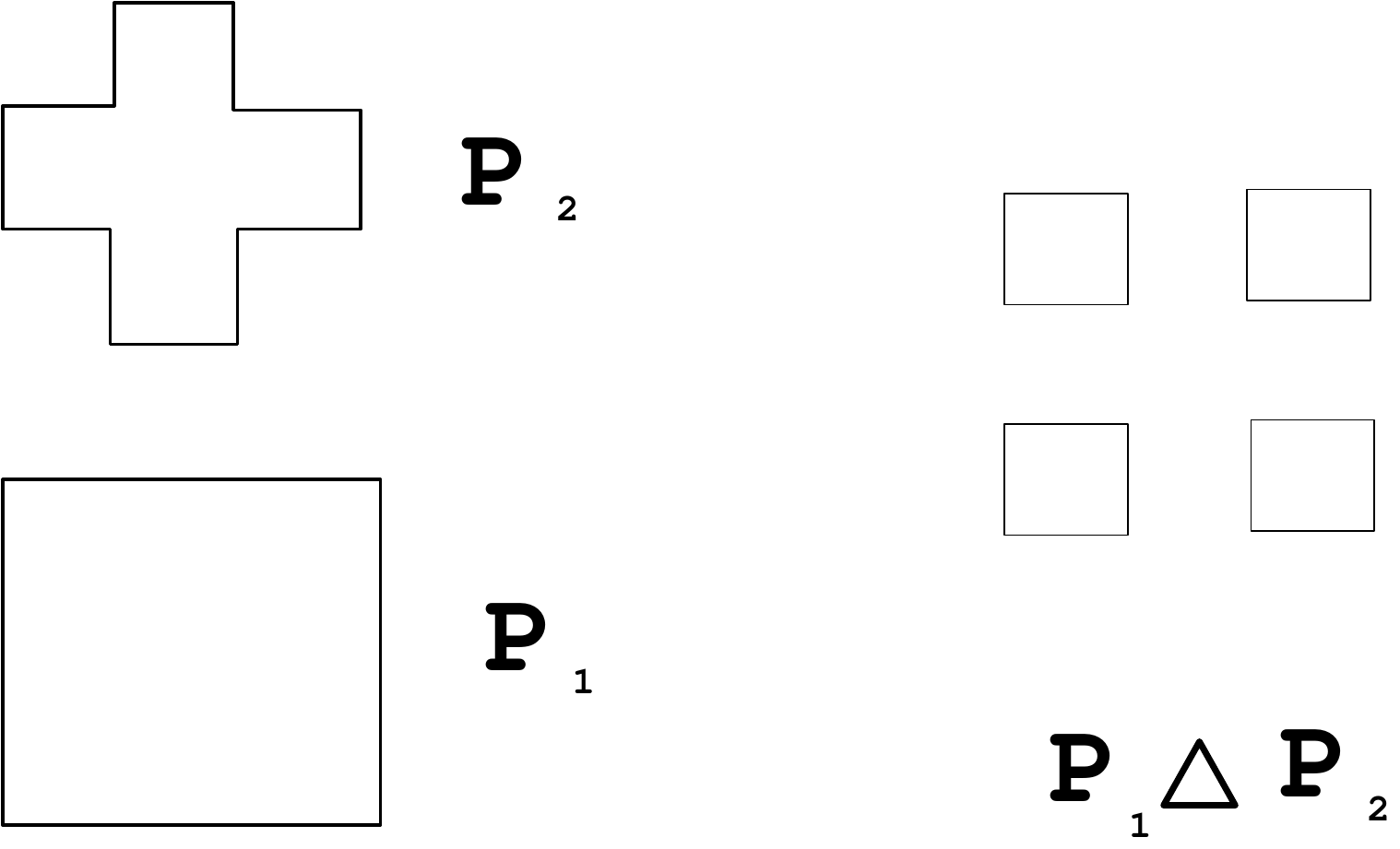}
\centering
\caption
{The energy of the surface M can be recovered by using images of the surface only on one
sequence of the parallel  planes (\ref{commonact}). For the surface on the picture we have
$k(P_1)=4,k(P_2)=12, l(\emptyset \bigtriangleup
P_1)=12, l(P_1 \bigtriangleup  P_2)=16,
l(P_2 \bigtriangleup\emptyset)=12$. Here $\emptyset$ denotes
 an empty polygon and $\bigtriangleup$ a symmetric difference (\ref{symmetricdifference}).
 Summing all these
quantities in accordance with the formula (\ref{commonact})
we shall get $A(M)= 4 + 12 +12+16+12 = 56$ times
$ a {\pi \over 2}$. This, as it should,  coincides with
the previous result on Fig.~\ref{fig3}. With the use of this representation the transfer
martix (\ref{tranmat}) can be viewed as describing the propagation of the polygon-loop
$P_{1}$ at time $\tau$ to another polygon-loop $P_{2}$ at
the time $\tau +1$.
 } \label{fig4}
\end{figure}

\section{\it  Dual Spin Systems }

Of a special  interest is a system with self-intersection coupling constant
equal to zero, $k=0$ \cite{8,9}. Because in that case the surfaces can freely self-intersect,
the system has even higher symmetry: one can flip spins $\sigma \rightarrow -\sigma$
on any set of planes orthogonal to the axis $x,y,z$.
In this limit the  Hamiltonian (\ref{spinsystem}) reduces to the following form:
\cite{8,9}
\be\label{kzerosystem}
H_{gonihedric}^{3d}=-\frac{1}{2}\sum_{\vec{r},
\vec{\alpha},\vec{\beta}} \sigma_{\vec{r}}
\sigma_{\vec{r}+\vec{\alpha}}
\sigma_{\vec{r}+\vec{\alpha}+\vec{\beta}}
\sigma_{\vec{r}+\vec{\beta}}
\ee
and  {\it its vacuum state degeneracy increases and is equal to $2^{3N}$
\cite{8,9}.
The last case is a sort of "supersymmetric" point in the space of
gonihedric Hamiltonians } (\ref{spinsystem}).

It is well known that the two-dimensional Ising model is a sef-dual system and that
three-dimensional Ising model is dual to the gauge spin system\cite{Wegner:1984qt}.
This duality was
an important fact allowing to find the exact solution of the Ising model in two dimensions.
We were able to construct dual systems for the gonihedric spin systems in three and
four dimensions \cite{9,10}. In three dimensions the dual spin system is of the
form\cite{9}
\be \label{dual}
H^{3d}_{dual} = -\sum_{\xi}
R^{\chi}(\xi) \cdot R^{\chi}(\xi + \chi) +R^{\eta}(\xi) \cdot
R^{\eta}(\xi + \eta)+ R^{\varsigma}(\xi) \cdot R^{\varsigma}(\xi +
\varsigma),
\ee
where $\chi$,~$\eta$~and $\varsigma$ are unit
vectors in the orthogonal directions of the dual lattice and
$R^{\chi}$, $R^{\eta}$ and $R^{\varsigma}$ are one-dimensional
irreducible representations of the group $Z_{2} \times Z_{2}$.

Similar construction can be performed in  four dimensions \cite{10}.
But, unlike the three-dimensional case, where we set the self-intersection
coupling constant $k$ to be equal to zero in (\ref{gonih}), here appears a complication.
Indeed, if
in four dimensions we set the self-intersection coupling
constant $k$ to be equal to zero then the  Hamiltonian (\ref{gonih}) has one and
three plaquette terms  and if we take $k=1$,
then it has one and two plaquette terms. In order to have even high symmetry one
can choose special weights on the intersections. Consider the case
when the intersection of six plaquettes contributes zero energy so that it
can be uniquely decomposed into three flat pairs of plaquettes.
The intersection of four plaquettes
yields zero energy in the cases when the plaquettes lie on two
flat planes and with the energy equal to $a \pi /2$
if a pair of plaquettes is left out of a plane.
Therefore a four-plaquette
intersection also uniquely decomposes into two flat planes in
the first case and into one flat plane and one "corner" in the
second case. For this choice of the self-intersection energies  the
Hamiltonian has the form \cite{10}
\be
H^{4d}_{gonihedric}=-\sum_{pairs~of~parallel~plaq}
(\sigma \sigma \sigma \sigma )_{P}^{~~||}
(\sigma \sigma \sigma \sigma )_{P} \label{4Dham},
\ee
where the summation is extended over all pairs of parallel
plaquettes in $3d$ cubes of the $4d$ lattice.
Here the Ising spins $\sigma_{\vec r,\vec r+\vec \alpha}$
are located in the center of the links
$(\vec r, \vec r + \vec \alpha)$ of the four-dimensional lattice.
One can check that the low temperature expansion
of the partition function of this system,
\be
Z(\beta)=\sum_{\{\sigma\}}e^{-\beta H^{4d}_{gonihedric} }=
 ~\sum_{\{M\}} e^{-\beta A(M)},
\label{4Z}
\ee
is obtained by summation over all closed
surfaces $\{ M \}$ with the weight $e^{-\beta A(M)}$, where
the linear action $A(M)$ is given by the number of non-flat pairs of
plaquettes of the closed surface $M$.

The details of the construction of the dual Hamiltonian can be found
in \cite{10}.  Here we shall present its final form \cite{10}:
\be
H^{4d}_{dual}= -\sum_{\xi}\sum_{\nu \neq \mu}
\Lambda_{\nu,\mu}(\xi)\cdot \Gamma(\xi,\xi + e_{\mu}) \cdot
\Lambda_{\mu,\nu}(\xi + e_{\mu}). \label{4Ddual}
\ee
It  is a spin
system of six Ising spins $\Lambda_{\mu,\nu}(\xi) = \Lambda_{\nu,\mu}(\xi)$,
$\mu \neq \nu = 1,2,3,4$  located on every $vertex$ $\xi$ of
the lattice and of one Ising spin  $\Gamma(\xi,\xi +e_{\mu})$
located on the center of every $link$ $(\xi,\xi +e_{\mu})$, where
 $e_{\mu}$ are the unite vectors along the four axis.

Both Hamiltonian (\ref{dual}) and (\ref{4Ddual}) look
differently, but it is possible to rederive (\ref{dual})
in the form which is similar to (\ref{4Ddual}).
For that let us
introduce three different Ising spins $\{\Lambda_{1},
\Lambda_{2},\Lambda_{3} \}$ in every vertex $\xi$, then
\be
H^{3d}_{dual} = \sum_{i\neq j\neq k} \Lambda_{j}(\xi) \Lambda_{k}(\xi)
\Lambda_{j}(\xi+e_{i}) \Lambda_{k}(\xi+e_{i}) \label{L3Ddual}.
\ee
As we shall see in the next section, this approach allows to construct the corresponding
transfer matrix, to prove that it describes the propagation
of closed loops and in a special case to find its spectrum.
This will present a unique exact solution of the tree-dimensional
statistical spin system.

\section{\it Transfer Matrices and Exact Solutions}

\begin{figure}
\includegraphics[width=6cm]{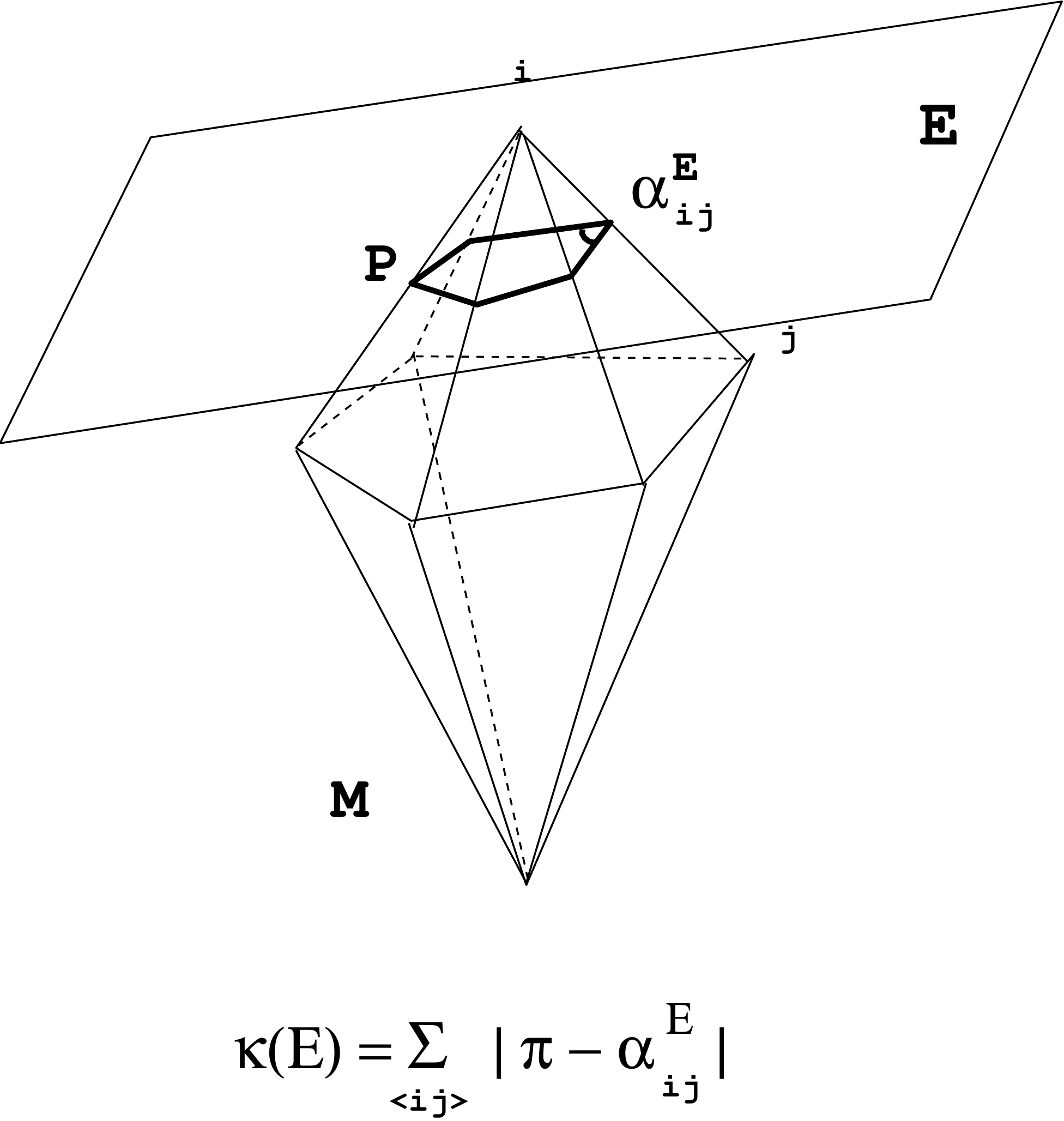}
\centering
\caption{ Intersection of the polyhedral surface $M$
by the plane $E$. The image of the surface on the plane is a
polygon $P$ and its curvature is equal to $k(E)$.
}
\label{fig5}
\end{figure}
In this section we  shall consider the above model of random
surfaces embedded into 3d Euclidean lattice $Z^3$. The reason to focus
on this particular case is motivated by the fact that one can
geometrically construct the corresponding transfer matrix  \cite{11}
and find its exact spectrum \cite{12,13,22}.

In order to find transfer matrix for this system we have to use the {\it Geometrical
theorem} proven in \cite{4}.
The geometrical theorem  provides  an equivalent
representation of the action $A(M)$ in terms of the absolute total curvature
$k(E)$ of the polygon $P(E)$ which appears in the intersection of the
two-dimensional plane $E$ with the given two-dimensional
surface $M$ (see Fig. \ref{fig5}):
\beqa
 k(E) =
\sum_{<i,j>} \vert \pi-\alpha^{E}_{ij} \vert  , \label{pol}
\eeqa
where $\alpha^{E}_{ij}$ are the angles of this polygon. They are
defined as the angles in the intersection of the two-dimensional
plane $E$ with the edge $<ij>$ (see Fig. \ref{fig5}).
The meaning of (\ref{pol}) is
that it measures the total revolution of the
tangent vectors to polygon $P_E$.
By integrating the total curvature $k(E)$ in (\ref{pol}) over all
intersecting planes $E$ we shall get the action $A(M)$ \cite{4}:
\be
A(M) = \frac{1}{2\pi}\int_{\{ E \} }k(E)dE  \label{curv}.
\ee
{\it Geometrical Theorem on a lattice \cite{7,8,9}.}
 One can find the same representation (\ref{curv}) for the action
$A(M)$ on a cubic lattice $Z^{3}$ by introducing a set of planes
$\{ E_{x} \}, \{ E_{y} \}, \{ E_{z} \} $ perpendicular to $x, y,z $ axis
on the dual lattice.
These planes will intersect a given surface $M$ and
on each of these planes we shall have an image of the  surface
$M$. Every such image is represented as a collection of
closed polygons $ Q(E)$ appearing in the intersection of
the plane with surface $M$ (see Fig. \ref{fig3},\ref{fig4}).
The energy of the surface $M$ is equal now to the
sum of the total curvature $k(E)$ of all these polygons
on  different planes:
\be
A(M) = \sum_{\{ E_{x},E_{y},E_{z}\} }
k( E ). \label{latcurv}
\ee
The total curvature  $k(E)$ is  the total number of polygon right angles.
 With (\ref{latcurv}) the partition function of the system (\ref{equalpartition}) can
be written in the form
\be\label{partfun}
Z(\beta) = \sum_{\{M\}}
\exp\{-2 \beta \sum_{\{E\}} k(E)\},
\ee
where the sum in the exponent can be represented as a product:
\be
\prod_{\{E\}}  e^{-2 \beta k(E)} =
\prod_{\{E_{z}\}}  e^{-2 \beta k(E_{z})}
\prod_{\{E_{y}\}}  e^{-2 \beta k(E_{y})}\prod_{\{E_{x}\}}
e^{-2 \beta k(E_{x})}. \label{product}
\ee
The goal is to express the energy functional (\ref{latcurv}) and the
product (\ref{product}) in terms of images on
parallel  planes in one fixed direction, let us say, $\{E_{z}\}$.
The question is: what kind
of information do we need to know on planes  $\{E_{z}\}$
in order to recover the values of the total curvature $k(E_{x})$
and $k(E_{y})$ on the planes  $\{ E_{y}\}$ and $\{E_{x}\}$?
The  contribution to the {\it curvature} $k(E_{x})+k(E_{y})$
of the polygons which are on the perpendicular planes between
$E^{i}_{z}$ and $E^{i+1}_{z}$  is equal to the {\it length of the polygons}
$Q_{i}$ and $Q_{i+1}$ without length of the common bonds:
\be\label{symmetricdifference}
l(Q_{i})+l(Q_{i+1})-2 \cdot l(Q_{i} \cap Q_{i+1} )
= l(Q_{i} \bigtriangleup  Q_{i+1}) ,  
\ee
where the polygon-loop~ $Q_{1} \bigtriangleup  Q_{2}~ \equiv~ Q_{1} \cup
Q_{2} ~\backslash~ Q_{1} \cap  Q_{2} $~ is a union
$Q_{1} \cup Q_{2}$ without intersection $Q_{1} \cap  Q_{2}$.
Therefore the  energy (\ref{latcurv}) can be expressed by using images
only on $\{E_{z}\}$ planes:
\be\label{commonact}
A(M) = \sum_{\{ E\}} k(E) = \sum_{\{ E_{z} \}}
k(Q_{i}) + l(Q_{i} \bigtriangleup  Q_{i+1}).
\ee
The partition function (\ref{partfun}) can be now represented in the form
\be\label{trace}
Z(\beta) =
\sum_{\{Q_{1},Q_{2},...,Q_{N}\}}~ K_{\beta}(Q_{1},Q_{2})\cdots
K_{\beta}(Q_{N},Q_{1})  = tr K^{N}_{\beta},
\ee
where $K_{\beta}(Q_{1},Q_{2})$ is the transfer matrix of  size
$\gamma \times \gamma $, defined as
\be\label{tranmat}
K (Q_{1},Q_{2}) =
\exp \{-\beta ~[k(Q_{1}) +
2 l(Q_{1} \bigtriangleup  Q_{2}) + k(Q_{2})]~ \},
\ee
where $Q_{1}$ and $Q_{2}$ are closed polygon-loops on a two-dimensional
lattice of size $N \times N$,  $k(Q)$ is the curvature and $l(Q)$ is
the length of the polygon-loop $Q$.   The total
number of polygon-loops is $\gamma = 2^{N^2}$.
The transfer matrix (\ref{tranmat}) can be viewed as
describing  the propagation of the polygon-loop $Q_{1}$ at the time $\tau$ to
another polygon-loop $Q_{2}$ at the time $\tau +1$.

The eigenvalues of the transfer matrix $K(Q_{1},Q_{2})$ define all
statistical properties of the system and can be found as a solution of the
following integral equation:
\be
\sum_{\{Q_{2}\} }K(Q_{1},Q_{2})~\Psi(Q_{2})=
\Lambda(\beta)~\Psi(Q_{1})  \label{inte}.
\ee
In the approximation when we drop the
curvature term in  $K(Q_{1},Q_{2})$
the spectrum can be evaluated exactly \cite{11,12,13}:
\be
 \Lambda_{P} = \sum_{\{Q\} } e^{-i\pi s(P \cap Q) - 2 \beta l(Q) },
\ee
and it identically coincides with the spin correlation functions of the 2d Ising model.
{\it In other words, the eigenvalues are expressed  in terms of all $\gamma = 2^{N^2}$
spin correlation functions $<\sigma_{i_1},...,\sigma_{i_n}>$, where spins are
located inside the  polygon P}.
Because the expression $\sum_{\{Q\} } e^{ - 2 \beta l(Q)} $ is the
partition function of the 2d Ising model,  we  see that
the largest eigenvalue $\Lambda_{0}$
is exactly equal to a corresponding partition function:
\be
Z(\beta) = Z_{Ising}(\beta).
\ee
It appears that the identical model was solved long ago in
\cite{22} and was called no-ceiling model (fuki-nuke model in Japanese). Recently a new development
took place in the articles \cite{23,24,25,16}, where anisotropic models were also 
considered.
\begin{figure}
\includegraphics[width=6cm]{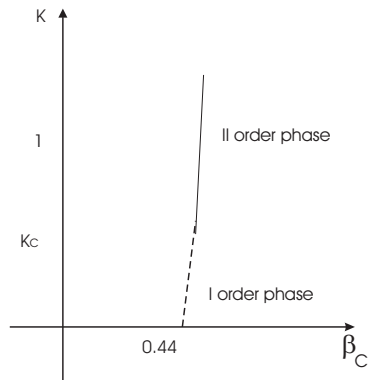}
\centering
\caption{The diagram shows different phases of the system (\ref{spinsystem})
as a function of the intersection coupling constant k. The partition function (\ref{equalpartition})
is  a function of temperature $\beta$ and of the coupling constant k.
 At $k=0$  the system (\ref{spinsystem}),
 (\ref{equalpartition})   reduces to
 (\ref{kzerosystem}), (\ref{partfun}) and (\ref{trace}) with transfer matrix (\ref{tranmat})
 and demonstrates, a strong first order phase transition.  This is
 the "supersymmetric" system defined in the  sections 3 and 4.  As the intersection coupling
 constant  k increases, the first order phase transition is weakening, and at $k_c \approx 0.5$
the system undergoes a second order phase transition at $\beta_c \approx 0.44$
 and has the critical indices ($\nu \approx 0.34$) which are different from those of the 3D Ising model
 ($\nu \approx 0.64$).
These systems belong to different classes of universality.  A similar behaviour
demonstrates the system (\ref{gonih}) in four dimensions. {\it The
 critical behaviour and the phase structure of  the gonihedric
spin systems essentially depend on the intersection coupling  constant k.} }
\label{fig6}
\end{figure}

\section{\it  Monte-Carlo Simulation of Gonihedric Systems in Various Dimensions}

The phase structure of the spin systems can be studied by
using Monte-Carlo simulations \cite{200,201,Savvidy:2015jva}. 
The random surfaces
with area action in four dimensions are defined through the one-plaquette
self-dual gauge invariant action\cite{Wegner:1984qt}.
The simulations indicate that the phase transition is of the first order
\cite{Creutz:1979kf}. The four-dimensional Ising model exhibits critical
behaviour with infinite correlation length and is supposed to be  equivalent to a
Higgs-like theory in the continuum limit
\cite{Wilson:1974,Wilson:1973jj,Cardy:1989da,Aizenman:1983bz}.

\begin{figure}
\includegraphics[width=7cm]{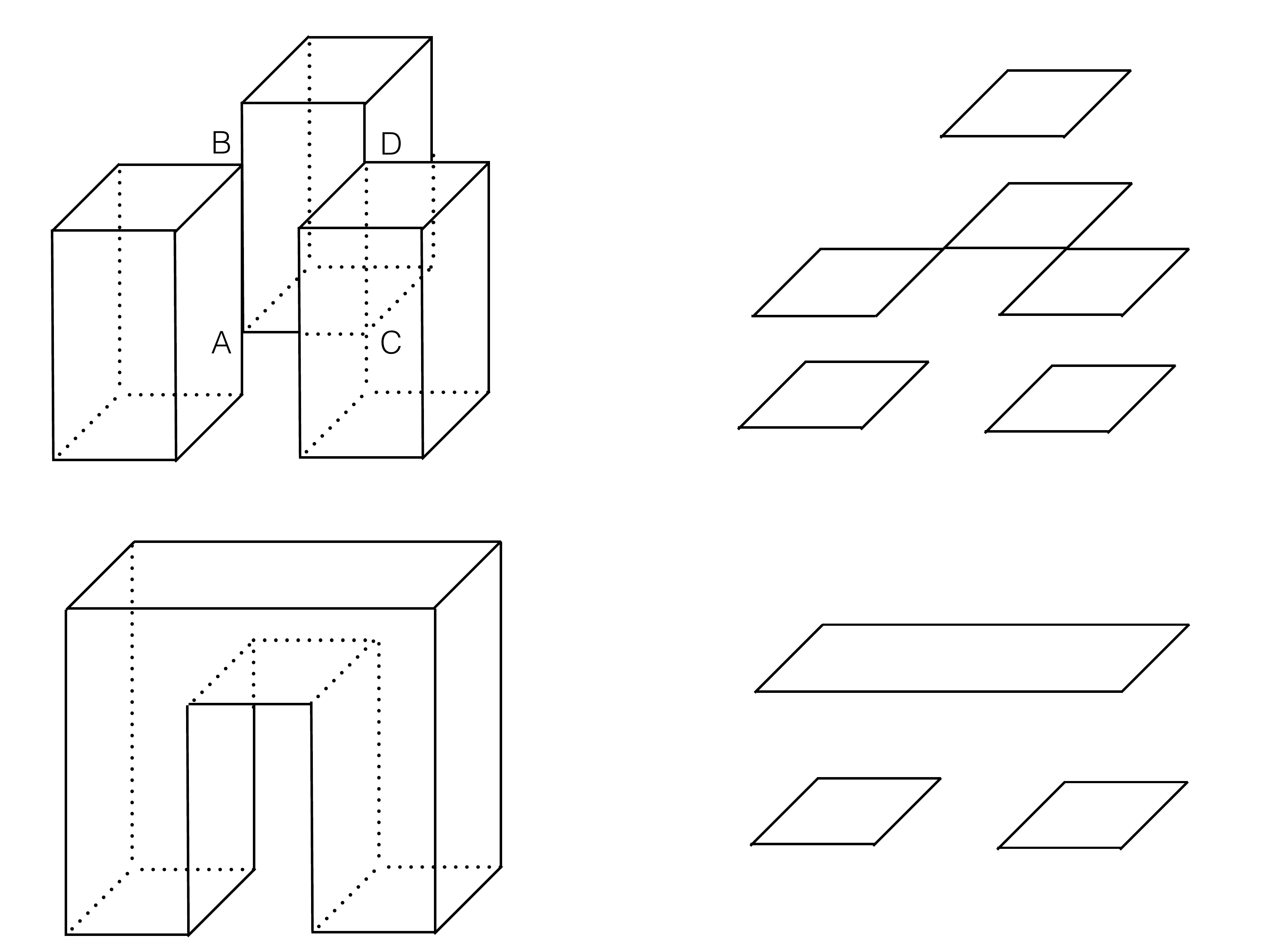}
\centering
\caption{The diagrams on the left show the main topological types of surfaces
with and without self-intersections.  On the
edges AB and CD take place the intersections of four plaquettes.
The figures on the right show the  horizontal  "time" slices of the
surfaces.  These time slices demonstrate that two  initial
strings propagate  in the vertical direction, then interact  and merge into the final
string.   The interaction "time" is proportional to the length of the edges AB and CD.
The intensity of string interactions is described by the  intersection coupling constant k.
In the limit $k \rightarrow \infty$  surfaces are self-avoiding. }
\label{fig7}
\end{figure}

The simulations of the gonihedric system (\ref{spinsystem}) in three dimensions  demonstrate
that it exhibits a second order phase transition \cite{14,15,16,29} (see Fig. \ref{fig6}).
In four dimensions the system (\ref{gonih}) also
undergoes  a second order phase transition\cite{17,Ambjorn:1998ff}, suggesting
that in the continuum limit there exists a string theory in four dimensions.
A further study of the critical properties of the proposed models can be rewarding,
specifically  {\it the scaling behaviour of the intersection lengths} $<n_4>$ and
$<n_6>$ \cite{17,Ambjorn:1998ff}.
These are the disorder parameters of the system ( they vanish in the low-temperature phase
and are non-zero at high-temperature phase ) defined as the derivative of the partition
function (\ref{equalpartition}) with respect to the coupling $k$. They represent
the  average length  $\mu_D$ of the self-intersection edges (see Fig. \ref{fig7})
\be
\mu_3 = - {1\over \beta}{\partial \ln Z \over \partial k} = 4 < n_4 >,~~~
\mu_4 = - {1\over \beta}{\partial \ln Z \over \partial k}=<4 \bar{n}_{4} 
+ 6 \bar{\bar{n}}_{4} + 12n_{6}>.
\ee
The second derivative with respect to $k$ defines the intersection susceptibility.
These order parameters are analogous to the magnetisation in the case of Ising model
and to the density in the case of liquid-gas transitions.

The other interesting property of the system is that the relaxation to the equilibrium
state is very slow, like in spin glasses \cite{10,17,26,29,30,32,33,34,Antenucci:2014uea,Johnston:2015vga,Mueller:2017znm,Mueller:2014yda,Mueller:2014oza,Mueller:2014noa,Mueller:2013ida}.
The reason is rooted in high symmetry of the system (\ref{kzerosystem}), its {\it energy states are exponentially degenerated} \cite{7,8,9}.

\section{\it  Random Manifolds with Gonihedric Energy }

Similar construction can be extended to the random manifolds of high dimension,
that is, for three-,  four- and higher-dimensional manifolds, sometimes  called
p-branes (p=2,3,..). We defined the corresponding energy functionals and transfer matrices, as well as equivalent spin systems\cite{18,19,20}. This allows to simulate random manifolds of
higher dimensionality on hypercubic lattices.

\section{\it Memory Devices Based on Gonihedric Systems }

{\it The gonihedric spin systems are the systems of very high symmetry, their  
states are  exponentially  degenerate. The discovery and the understanding of these symmetry in} \cite{9,10} {\it was essential ingredient in the construction of the dual systems}. The rate of degeneracy of the vacuum state depends on the self-intersection coupling constant $k$ \cite{9,10}.

{\it If $k \neq 0$, one can flip spins on any set of parallel
planes and  the degeneracy of the vacuum state is equal to $3\cdot 2^N$ for
the lattice of the size $N^3$. The system with the self-intersection coupling constant $k=0$ 
is even more symmetric, one can flip spins $\sigma \rightarrow -\sigma$
on any set of planes orthogonal to the axis $x,y,z$ and is equal to} $2^{3N}$ \cite{9,10}.  

These vacuum states,  are separated by the potential barriers, therefore one can suggest to use such systems
as storages of the binary information \cite{21,25}. Because there is no interface energy
proportional to the area in these systems one can store one bit of information in a very small
region of the gonihedric "crystal". It is an interesting and challenging problem to
construct an artificial material which will have a corresponding structure.

\section*{\it Acknowledgments}
This project has received funding from the European Union's Horizon 2020 research and innovation programme under the Marie Sk\'lodowska-Curie grant agreement No 644121.

\end{document}